\documentclass[twocolumn]{aastex631}

\usepackage{amsmath,soul}

\newcommand{\mii}{Mg\,\textsc{ii}}
\newcommand{\civ}{C\,\textsc{iv}}

\begin{document}

\title{Effects of type Ia supernovae absolute magnitude priors on the Hubble constant value}

\author[0000-0001-8919-7409]{Yun Chen}
\affiliation{Key Laboratory for Computational Astrophysics, National Astronomical Observatories, Chinese Academy of Sciences, Beijing 100101, China}
\affiliation{College of Astronomy and Space Sciences, University of Chinese Academy of Sciences, Beijing, 100049, China}

\author[0000-0002-1638-2264]{Suresh Kumar}
\affiliation{Department of Mathematics, Indira Gandhi University, Meerpur, Haryana 122502, India}

\author[0000-0002-7307-0726]{Bharat Ratra}
\affiliation{Department of Physics, Kansas State University, 116 Cardwell Hall, Manhattan, KS 66506, USA}

\author[0000-0002-9855-2342]{Tengpeng Xu}
\affiliation{Key Laboratory for Computational Astrophysics, National Astronomical Observatories, Chinese Academy of Sciences, Beijing 100101, China}
\affiliation{College of Astronomy and Space Sciences, University of Chinese Academy of Sciences, Beijing, 100049, China}

\email{chenyun@bao.ac.cn}
\email{suresh.math@igu.ac.in}




\begin{abstract}

We systematically explore the influence of the prior of the peak absolute magnitude ($M$) of type Ia supernovae (SNe Ia) on the measurement of the Hubble constant ($H_0$) from SNe Ia observations. We consider five different data-motivated $M$ priors, representing varying levels of dispersion, and assume the spatially-flat $\Lambda$CDM cosmological model. Different $M$ priors lead to relative changes in the mean values of $H_0$ from 2\% to 7\%. Loose priors on $M$ yield $H_0$ estimates consistent with both the Planck 2018 result and the SH0ES result at the 68\% confidence level. We also examine the potential impact of peculiar velocity subtraction on the value of $H_0$, and show that it is insignificant for the SNe Ia observations with redshift $z > 0.01$ used in our analyses. We also repeat the analysis in the cosmography model and find very similar results. This suggests that our results are robust and model independent.

\end{abstract}



\section{Introduction} \label{sec:intro}

The Hubble constant, $H_0$, that is the current value of the cosmological expansion rate, is inversely proportional to the age of the universe, $t_0\propto1/H_0$, and is a key cosmological parameter, see e.g., \citet{2010ARA&A..48..673F}. In addition, when estimating cosmological model parameters from observational data, $H_0$ can be degenerate with other cosmological parameters such as the nonrelativistic matter density parameter or the spatial curvature density parameter, and so the value of $H_0$  can affect observational constraints on other parameters \citep[see e.g.][] {2016PhLB..752...66C, 2016ApJ...829...61C, 2017ApJ...835...86C, ParkRatra2020, CaoRyanRatra2020, KhadkaRatra2020, 2021MNRAS.503.2179Q}. 
 
Measuring $H_0$ is challenging, especially because of the difficulty in determining distances to astronomical objects \citep{2021A&ARv..29....9S}. Many methods have been used to improve the accuracy and precision in the measurement of $H_0$ since Hubble's first measurement in 1929. Nevertheless, there have been many differing estimates of $H_0$ over the following more than half a century until measured values began to converge at the start of the 21st century, led by the 2001 measurement of  $H_0 = 72\pm 8$ km s$^{-1}$ Mpc$^{-1}$ (1$\sigma$ error including systematics) from the Hubble Space Telescope (HST) Key Project \citep{2001ApJ...553...47F}. 

In the two decades since, $H_0$ measurements with smaller error bars have become available. For example, the Planck satellite 2013 results included an estimate of $H_0=67.9\pm1.5$ km s$^{-1}$ Mpc$^{-1}$ \citep{2014A&A...571A..16P} inferred from measurements of (higher-redshift) cosmic microwave background (CMB) temperature and lensing-potential power spectra in the framework of the standard spatially-flat $\Lambda$CDM cosmological model \citep{Peebles1984}. Slightly earlier, \citet{2011ApJ...730..119R} estimated $H_0=73.8\pm2.4$ km s$^{-1}$ Mpc$^{-1}$ from (lower-redshift) apparent magnitude versus redshift measurements of SNe Ia that were calibrated using HST observations of Cepheid variables in host galaxies of eight nearby SNe Ia. Statistically, the difference between these two results is $2.1\sigma$. In the same year, from a median statistics analysis of Huchra's compilation of 553 measurements of $H_0$, \citet{2011PASP..123.1127C} found $H_0=68\pm2.8$ km s$^{-1}$ Mpc$^{-1}$, at $1\sigma$ and including systematic errors \citep[also see][]{2001ApJ...549....1G, ChenGottRatra2003, Calabreseetal2012}, $1.6\sigma$ lower than the Cepheid-calibrated SNe Ia estimate \citep{2011ApJ...730..119R}. 

More recently, the difference between some local and  high-redshift determinations of $H_0$ has increased to $\sim 5\sigma$, for instance, $H_0 = 73.04\pm1.04$ km s$^{-1}$ Mpc$^{-1}$ from the Supernovae and H0 for the Equation of State of dark energy (SH0ES) project that use (lower-redshift) Cepheid-calibrated SNe Ia data \citep{2022ApJ...934L...7R}, and $H_0 = 67.36\pm0.54$ km s$^{-1}$ Mpc$^{-1}$ from (higher-redshift) Planck 2018 CMB observations under the standard $\Lambda$CDM model \citep{2020A&A...641A...6P}. There are however other local $H_0$ estimates that are not as high, for instance, $H_0 = 69.8 \pm 1.7$ km s$^{-1}$ Mpc$^{-1}$ from SNe Ia calibrated using tip of the red-giant branch data \citep{Freedman2021} or the flat $\Lambda$CDM model value provided in \cite{2023PhRvD.107j3521C}, $H_0=69.5\pm 2.4$ km s$^{-1}$ Mpc$^{-1}$, from a joint analysis of Hubble parameter, baryon acoustic oscillation (BAO), Pantheon+ SNe Ia, quasar angular size, reverberation-measured \mii\ and \civ\ quasar, and 118 Amati correlation gamma-ray burst data (that is independent of CMB data).

The difference between the higher-redshift (Planck and other CMB) $H_0$ values and some of the lower-redshift $H_0$ values is known as the Hubble tension. A number of schemes have been proposed to explain or resolve the $H_0$ tension. In general, these fall into two categories \citep{2021A&ARv..29....9S, 2021CQGra..38o3001D,Perivolaropoulos:2021jda, Abdalla:2022yfr, HuWang2023, Vagnozzi:2023nrq, Freedman:2023jcz, Wangetal2024,2021ApJ...912..150D,2023MNRAS.521.3909B}: (i) the difference is due to unresolved or unknown systematics in the analyses of the higher-redshift CMB data and/or some of the lower-redshift local distance-ladder data; or (ii) the difference indicates that the standard flat $\Lambda$CDM model is an inadequate cosmological model. The second category includes a variety of proposed cosmological models, some with modified, relative to flat $\Lambda$CDM, low-redshift or late-time evolution, e.g., some modified  gravity theories \citep{2020PhRvD.101j3505D}; some specific dark energy models \citep{2020PhRvD.102l1302D, 2021PhRvD.103l1302C,2022PhLB..83137174B,2021PhRvD.103j3509K};  models with a significant local inhomogeneity \citep{2013PhRvL.110x1305M, 2019ApJ...875..145K,2024arXiv240114170H}; and early dark energy models with different high-redshift or early-time evolution \citep{2019PhRvL.122v1301P,2021PhRvD.103d3518T,2019PhRvD.100j3524R,2020PhRvD.102b3529B,2021PhRvD.103b3530A}; as well as models that include dark matter-neutrino interactions \citep{2018PhRvD..97d3513D}, or neutrino self-interactions \citep{2020PhRvD.101l3505K}; models that include dark radiation \citep{2023arXiv230709863L, 2023arXiv230615067G}; and models that rapidly transition from an anti-de Sitter vacuum to a de Sitter vacuum at redshift $z\sim 2$ \citep{Akarsu:2019hmw,Akarsu:2021fol,Akarsu:2022typ,Akarsu:2023mfb}. 

Although the proposed alternate cosmologies are valuable in the context of the $H_0$ tension \citep{2019ApJ...886L..23L,  2021ApJ...915..123S, 2023A&A...674A..45J}, there also are a class of solutions that indicate that prior knowledge of the peak absolute magnitude $M$ of SNe Ia could have a non-ignorable influence on the value of $H_0$ determined from SNe Ia observations. For example, \citet{2023MNRAS.520.5110P} show that introducing a single new degree of freedom in the cosmological analysis of the Pantheon+ SNe Ia compilation can change the best-fitting value of $H_0$; here instead of the usual single peak absolute magnitude parameter $M$, they allow for a change of the absolute magnitude at a transition distance $d_{\rm crit}$, i.e., $M = M_{<}$ when distance $d<d_{\rm crit}$, and $M = M_{>}$ when $d > d_{\rm crit}$. Also, some studies claim that intrinsic dispersion or evolution of $M$ can sufficiently alter the estimate of $H_0$ from SNe Ia data \citep{2017A&A...602A..73T, 2019A&A...625A..15T, 2020JCAP...07..045D}. 

While the correlation between $H_0$ and $M$ is well known and widely discussed in the literature, previous research has not systematically investigated the impact of the prior of $M$ on the measurement of $H_0$ from SNe Ia data. Here we try to  systematically assess how the $M$ prior affects the determination of $H_0$ from SNe Ia data, by considering five typical data-motivated priors for $M$ that have previously been discussed in the literature. 

The rest of the paper is organized as follows. In Section \ref{sec:method}, we summarize the methodology of using the recent Pantheon+ SNe Ia compilation in cosmological analyses, and discuss how two potential factors may affect the constraint on $H_0$ in such analyses. We present and discuss our results in Section \ref{sec:results}. In the last section, we summarize our main findings.

\section{Data and Methodology} \label{sec:method}

\subsection{Cosmological application of SNe Ia}

SNe Ia apparent magnitude versus redshift measurements provided the first convincing evidence for currently accelerating cosmological expansion \citep{1998AJ....116.1009R, 1999ApJ...517..565P}. Over the past quarter century, many more SNe Ia have been discovered, and with improved data reduction techniques this has resulted in refinements of this neoclassical cosmological test \citep{2010A&A...523A...7G, 2014A&A...568A..22B, 2018ApJ...859..101S, 2019ApJ...872L..30A, 2020RAA....20..151G, 2022MNRAS.514.5159M, Rubinetal2023, DES2024}. In this work, we use the recent Pantheon+ sample that includes 1701 light curves of 1550 distinct SNe Ia in the redshift range $0.001 < z < 2.26$ \citep{2022ApJ...938..110B}.

The Pantheon+ sample is compiled from 18 different surveys, but with the SNe Ia light curves now uniformly fit to SALT2 model light curves \citep{2010A&A...523A...7G, 2022ApJ...938..111B}. The SALT2 light-curve fit returns four parameters for each supernova: i) the light-curve amplitude $x_0$, which is related to the apparent B-band peak magnitude $m_B \equiv -2.5\log_{10}(x_0)$; ii) the stretch parameter $x_1$, which  corresponds to the light-curve width; iii) the light-curve color parameter $c$, which has contributions from both intrinsic color and dust; and, iv) $t_0$, which denotes the time of peak brightness. 

After fitting the light curves of SNe Ia in the Pantheon+ sample with the SALT2 model, \citet{2022ApJ...938..110B} determined the Hubble diagram and the corresponding covariance matrices by using the “BEAMS with Bias Corrections” (BBC) method \citep{2017ApJ...836...56K}, where BEAMS (Bayesian Estimation Applied to Multiple Species) provides a unified classification and parameter estimation methodology \citep{2007PhRvD..75j3508K}. In practice, the corrected apparent magnitude is used as an observable quantity for each Pantheon+ SN Ia \citep[see Table 7 of][]{2022ApJ...938..110B}, i.e.,
	\begin{equation}
		\begin{aligned}
			m^{\mathrm{corr}}_B&\equiv\mu+M \\
			&=m_B+(\alpha x_1-\beta c-\delta_{\mathrm{bias}}+\delta_{\mathrm{host}}),
		\end{aligned}
  \label{eq:m_obs}
	\end{equation}
where $\mu$ is the distance modulus and $M$ is the peak absolute B-band magnitude. Here $\alpha$ and $\beta$ are global nuisance parameters relating stretch and color to luminosity,  $\delta_{\mathrm{bias}}$ is a correction term related to selection biases, and  $\delta_{\mathrm{host}}$ is a correction term originating from the residual correlation between the standardized brightness of the supernova and the host galaxy mass. 

In a given cosmological model, the predicted apparent magnitude at redshift $z$ can be computed from 
	\begin{equation}
		m_{\mathrm{th}}(z;\textbf{p})=\mu_{\mathrm{th}}(z;\textbf{p})+M,
		\label{eq:m_th}
	\end{equation}
where $\textbf{p}$ is the set of cosmological model parameters and the predicted distance modulus is
	\begin{equation}
		\mu_{\mathrm{th}}(z;\textbf{p}) = 5\log_{10}\left[\frac{d_{\mathrm L}(z;\textbf{p})}{\mathrm{Mpc}}\right]+25.
  \label{eq:mu_th}
	\end{equation}
Here $d_{\mathrm{L}}(z;\textbf{p})$ is the predicted model-based luminosity distance which is related to the Hubble parameter $H(z)$.  

In our analysis here we use the standard spatially-flat $\Lambda$CDM cosmological model in which the luminosity distance can be expressed as
\begin{equation}
d_L(z;\textbf{p}) = c_0 (1+z)\int_{0}^{z}\frac{dz'}{H(z';\textbf{p})},
 \label{eq:dl_th}
\end{equation}
where $c_0$ is the speed of light. In the flat $\Lambda$CDM model the expansion rate or Hubble parameter is 
\begin{equation}
H(z;\textbf{p}) = H_0\sqrt{\Omega_{\rm m0}(1+z)^3+(1-\Omega_{\rm m0})},
\label{eq:Hz}
\end{equation}
where the model parameter set $\textbf{p} = (\Omega_{\rm m0},H_0)$ consists of the current value of the non-relativistic matter density parameter $\Omega_{\rm m0}$ and $H_0 = H(z=0)$.

In our analyses, the likelihood of the SNe Ia sample is computed using
\begin{equation}
\mathcal{L}(\mathbf{p})\propto e^{-\chi^2/2},
\label{eq:likelihood}
\end{equation}
where $\chi^2$ is
\begin{equation}
\chi^2 = \Delta\overrightarrow{m}^T\mathbf{C}^{-1} \Delta\overrightarrow{m},
\label{eq:chi2}
\end{equation}
with the residual vector $\Delta\overrightarrow{m}_i=m^{\mathrm{corr}}_{B,i}-m_{\mathrm{th}}(z_i;\textbf{p})$. Here $m^{\mathrm{corr}}_{B,i}$ and $m_{\mathrm{th}}(z_i;\textbf{p})$ for the $i^{\mathrm{th}}$ supernova can be computed from equations (\ref{eq:m_obs}) and (\ref{eq:m_th}), respectively. The covariance matrix $\mathbf{C}$ includes contributions from both statistical and systematic errors and is available on the Pantheon+ website.
\footnote{https://github.com/PantheonPlusSH0ES/DataRelease}

Following \citet{2019PhRvD.100b3532A,2023PDU....3901162A}, we use the PyMultiNest code \citep{2014A&A...564A.125B} to compute posterior probability distributions for the model parameters, where the likelihood is defined in equation (\ref{eq:likelihood}). PyMultiNest is a Python interface for the MultiNest algorithm \citep{2009MNRAS.398.1601F}, which is  a generic Bayesian inference tool that uses nested sampling \citep{2004AIPC..735..395S} to obtain both posterior samples and Bayesian evidence estimates.

\subsection{Effects of assumed absolute magnitudes and truncation redshifts}

The degeneracy between the peak absolute magnitude $M$ and the Hubble constant $H_0$ is a well-known issue in cosmological analyses with SNe Ia data, and how to treat this degeneracy --- besides the traditional approach of using local calibrators to break it --- has also been discussed in the literature \citep{2021MNRAS.504.5164C, 2023arXiv230702434C, 2021MNRAS.505.3866E, 2022Univ....8..502P, 2023MNRAS.520.5110P, 2023IJAA...13...39M}. Here we use five different data-motivated priors on $M$ to break this degeneracy. 

In addition, the relative uncertainties of the cosmological redshifts arising from peculiar velocity corrections are larger for low-$z$ SNe Ia than for those in the high-$z$ realm. To assess the effect of peculiar velocity subtraction, we choose to remove from the analysis low-$z$ SNe Ia at different redshift thresholds.

\subsubsection{Priors on the peak absolute magnitude of SNe Ia }
\label{sec:prior}

To explore the effects of the prior on $M$ in the cosmological application of SNe Ia we consider five different priors, that have been discussed in the literature, which  we label as \textbf{Prior I--V}.

\begin{itemize}
\item \textbf{Prior I: A Gaussian prior from the Cepheid-calibrated distance ladder method.}
\citet{2022ApJ...934L...7R} obtain $M = -19.253\pm0.027$ by using the Cepheid calibration of SNe Ia based on observations of the SH0ES program. As discussed in \citet{2021MNRAS.505.3866E}, when combining SH0ES data with other astrophysical data to constrain late-time physics, one should impose a SH0ES prior on $M$ and not on $H_0$.
\item \textbf{Prior II: A Gaussian prior from the BAO-calibrated inverse distance ladder method.}
By using Gaussian process regression, \citet{2023PhRvD.107f3513D} obtain $M = -19.396\pm0.016$. Here Gaussian process cosmological reconstruction use Pantheon SNe Ia and BAO data. Since higher-$z$ BAO data are used, this process is sometimes called the ``inverse distance ladder'' method, to distinguish it from the traditional distance ladder method which uses nearby calibrators such as Cepheid variables.
\item \textbf{Prior III:  A Gaussian prior from the Multicolor Light Curve Shape method.}
\citet{2000ApJ...536..531W} finds $M=-19.33\pm0.25$ from observations of 18 SNe Ia with distance moduli obtained using the Multicolor Light Curve Shape (MLCS) method \citep{1998AJ....116.1009R}. 
\item \textbf{Prior IV: A loose top-hat prior.}
We also consider a loose top-hat prior with  $-20.0 < M < -18.0$, which is broad enough to cover the corresponding values for almost all known SNe Ia \citep{2002AJ....123..745R, 2011MNRAS.412.1441L, 2016MNRAS.460.3529A}.
\item \textbf{Prior V: A much looser top-hat prior.}
To examine how the width of a top-hat prior on $M$ impacts the constraint on $H_0$, we also consider a much looser top-hat prior compared to \textbf{Prior IV}, i.e., $-21.0 < M < -17.0$.
\end{itemize}

It is believed that the formation mechanism of SNe Ia implies that they possess an almost identical peak absolute magnitude \citep{1979ApJ...232..404C, 1995PhR...256...53B}, however, the diversity of progenitors of SNe Ia may result in a dispersion in $M$ among the SNe Ia \citep{2011NatCo...2..350H, 2018PhR...736....1L, 2018RAA....18...49W,2023ApJ...953...13F}.  Additionally, observationally, absolute magnitudes of SNe Ia depend on properties of their host galaxies \citep{1993ApJ...413L.105P, 2010MNRAS.406..782S, 2010ApJ...715..743K, 2011PASP..123..230K, 2016MNRAS.460.3529A, 2020ApJ...889....8K, 2020A&A...644A.176R, 2023MNRAS.518.1985M}. Therefore, an intrinsic dispersion of $M$ among the SNe Ia should be taken into account. The five priors discussed above represent five different levels of dispersion on $M$. Among them, \textbf{Priors I--III} are Gaussian priors with different levels of uncertainties, with the first two having $\sim 0.1$\% uncertainties, and the third a $\sim 1$\% uncertainty. 

\subsubsection{Truncation redshifts}
\label{sec:truncated_redshift}

As noted above, the relative uncertainties of the cosmological redshifts for the SNe Ia arising from peculiar velocity subtraction are larger in the low-redshift universe than in the high-redshift realm \citep{2011ApJ...741...67D, 2011PhRvD..83d3004B, 2020ApJ...904L..28H}. As discussed in \citet{2022ApJ...938..110B}, the Pantheon+ collaboration uses the $z > 0.01$ part of the sample in cosmological analysis to avoid dependence on the peculiar velocity correction model. To study potential effects of this correction, we study five different redshift-trucated SNe Ia samples, removing from the complete Pantheon+ compilation low-redshift SNe Ia at five different redshift thresholds, i.e., at $z > 0.01$, $z > 0.02$, $z > 0.05$, $z > 0.1$ and $z > 0.2$.

\section{Results} 
\label{sec:results}

\begin{table*}[htbp]
	
\caption{Observational constraints (mean values with 68\% confidence level errors) on the parameters $\Omega_{\rm m0}$ and $H_0$ ($\Lambda$CDM model as well as model-independent cosmography approach) obtained from the five redshift-truncated samples, for the five different priors of $M$. $H_0$ has units of km s$^{-1}$ Mpc$^{-1}$. Log-Bayesian evidence ($\ln B_i$) and relative log-Bayesian evidence ($\ln B_{i0}=\ln B_i-\ln B_0$) are also displayed in each case.} 
\label{tab:results}
\scalebox{0.6}{
\begin{tabular}{cllllrrllllllrr} 
\hline\hline
Data Sample	 & $M$ prior & & &$\Lambda$CDM model& & &&&& &Cosmography & approach&  &\\
 \cline{4-7}\cline{10-15}
 &&&$\Omega_{\rm m0}$ & $H_0$&$\ln{B_i}$&$\ln{B_{i0}}$&&&$H_0$&$q_0$&$j_0$&$s_0$&$\ln{B_i}$&$\ln{B_{i0}}$\\
 \hline
   & $-19.253\pm0.027$ & &$0.350\pm{0.017}$ & $72.84\pm0.81$& -974.83& 0.00 & && $72.84\pm0.85$&$-0.488\pm0.043$&$1.01\pm0.30$&$-0.11\pm0.72$ &-977.19&0.00\\
truncated sample& $-19.396\pm0.016$ & &$ 0.352\pm{0.017}$& $68.09\pm0.50$&-975.47&-0.64&& &$68.21\pm0.50$&$-0.490\pm0.444$&$1.03\pm0.30$&$ -0.05\pm0.73$&-977.43&-0.24\\ 
with $z>0.01$ & $-19.33\pm0.25$ & &$0.351\pm0.017$ & $71.58\pm6.24$ & -975.37 & -0.54 && &$70.90\pm5.32$&$-0.490\pm0.044$&$1.03\pm0.30$&$-0.65\pm0.72$& -977.73&-0.54\\
$(N = 1590)$   & $[-20.0, -18.0]$ & &$0.350\pm{0.018}$ & $71.00\pm11.35$ & -975.86 & -1.03 & &&$70.72\pm11.07$&$-0.492\pm0.044$&$1.04\pm 0.30$&$-0.51\pm0.72$&-977.75&-0.56\\
 & $[-21.0, -17.0]$ && $0.351\pm{0.017}$ & $70.56\pm11.74$ & -976.57& -1.74 & &&$71.18\pm11.26$&$-0.488\pm0.046$&$1.04\pm0.31$&$ -0.06\pm0.73$&-978.70&-1.51\\
    \hline
     & $-19.253\pm0.027$ && $0.358\pm0.017$ & $72.86\pm0.83$ & -855.80& 0.00 & &&$72.89\pm0.79$&$-0.474\pm0.048$&$1.00\pm0.33$&$-0.19\pm0.75$&-858.43&0.00\\
truncated sample & $-19.396\pm0.016$ & &$0.359\pm{0.017}$ & $68.24\pm0.47$ & -856.33& -0.53&& &$68.23\pm0.50$&$-0.476\pm0.047$&$1.01\pm0.32$&$-0.15\pm0.74$&-858.53&-0.10\\ 
 with $z>0.02$& $-19.33\pm0.25$ & &$0.359\pm0.017$ & $71.01\pm6.23$ & -856.34& -0.54& &&$70.37\pm5.21$&$-0.475\pm0.045$&$1.01\pm0.31$&$-0.18\pm0.73$&-858.94&-0.51\\
$(N = 1436)$   & $[-20.0, -18.0]$ & &$0.359\pm{0.017}$ & $71.89\pm10.93$ & -856.89 & -1.09 & &&$71.39\pm11.09$&$ -0.474\pm0.048$&$1.00\pm0.33$&$-0.17\pm0.76$&-859.05&-0.62\\
    & $[-21.0, -17.0]$ && $0.359\pm{0.017}$ & $70.39\pm11.79$ & -857.16 &-1.36 & &&$71.21\pm11.87$&$-0.477\pm 0.046$&$1.03\pm0.31$&$-0.12\pm0.73$&-859.63&-1.20\\
    \hline
    & $-19.253\pm0.027$ && $0.364\pm 0.021$ & $72.58\pm0.89$ & -617.92 &0.00& &&$72.61\pm0.93$&$-0.451\pm0.605$&$0.90\pm0.36$&$-0.39\pm0.79$&-620.67&0.00\\
truncated sample & $-19.396\pm0.016$ & &$0.364\pm0.020$ & $67.96\pm0.55$ & -618.36 & -0.44 && &$68.03\pm 0.58$&$-0.452\pm0.060$&$0.90\pm0.36$&$ -0.37\pm0.77$&-620.72&-0.05\\
with $z>0.05$ & $-19.33\pm0.25$ & &$0.364\pm{0.020}$ & $71.39\pm6.07$ & -618.28& -0.36 && &$71.33\pm5.65$&$-0.451\pm0.058$&$0.90\pm0.35$&$-0.38\pm0.78$&-620.84&-0.17\\
$(N = 1056)$   & $[-20.0, -18.0]$ && $0.364\pm0.020$ & $70.42\pm11.40$ & -618.89& -0.97 & &&$70.75\pm11.37$&$-0.456\pm0.058$&$0.94\pm0.36$&$-0.30\pm0.79$& -621.17&-0.50\\
    & $[-21.0, -17.0]$ && $0.364\pm0.021$ & $70.69\pm11.67$ & -619.43&-1.51 & &&$70.09\pm12.05$&$-0.451\pm0.060$&$0.89\pm0.36$&$-0.40\pm0.78$&-621.84&-1.17\\
    \hline
    & $-19.253\pm0.027$ && $0.373\pm0.024$ & $72.43\pm0.87$ & -555.71& 0.00 & &&$72.41\pm0.94$&$-0.422\pm0.072$&$0.80\pm0.39$&$-0.60\pm0.79$&-558.15&0.00\\
truncated sample & $-19.396\pm0.016$ & &$0.374\pm{0.024}$ & $67.77\pm0.60$ & -555.85& -0.14 && &$67.80\pm0.66$&$-0.433\pm0.071$&$0.87\pm0.38$&$-0.45\pm0.81$&-558.30&-0.15\\ 
with $z>0.1$ & $-19.33\pm0.25$ & &$0.371\pm0.024$ & $70.97\pm6.61$ & -555.83& -0.12 && &$70.30\pm5.24$&$-0.429\pm0.070$&$0.86\pm0.39$&$-0.48\pm0.79$&-558.33&-0.18\\
$(N = 960)$  & $[-20.0, -18.0]$ & &$0.373\pm0.024$ & $70.29\pm11.15$ & -556.43& -0.72 & &&$70.52\pm11.13$&$-0.427\pm0.070$&$0.84\pm0.38$&$-0.52\pm0.79$&-558.73&-0.58\\
    & $[-21.0, -17.0]$ && $0.374\pm0.025$ & $70.89\pm11.32$ &-557.07&-1.36& & &$70.72\pm11.75$&$-0.427\pm0.073$&$0.83\pm0.40$&$-0.52\pm0.82$&-558.76&-0.61\\
\hline
 & $-19.253\pm0.027$ && $0.369\pm0.028$ & $72.49\pm1.02$ & -414.51&0.00 & &&$72.46\pm1.02$&$-0.430\pm0.831$&$0.83\pm0.40$&$-0.48\pm0.82$&-417.34&0.00\\
truncated sample  & $-19.396\pm0.016$ & &$0.369\pm{0.029}$ & $67.82\pm0.68$ & -414.83& -0.32&& &$67.83\pm0.72$&$-0.437\pm0.078$&$0.87\pm0.39$&$-0.39\pm0.81$&-417.42&-0.08\\ 
with $z>0.2$ & $-19.33\pm0.25$ & &$0.370\pm{0.029}$& $70.83\pm6.37$ & -414.96 & -0.45 && &$70.44\pm5.74$&$-0.435\pm0.084$&$0.86\pm0.39$&$-0.42\pm0.82$&-417.44&-0.10\\
$(N = 753)$   & $[-20.0, -18.0]$ & &$0.370\pm0.028$ & $70.49\pm11.17$& $-415.25$& -0.74&&& $71.11\pm11.54$&$-0.428\pm0.082$&$0.84\pm0.39$&$-0.48\pm0.82$&-417.89&-0.55\\
    & $[-21.0, -17.0]$ && $0.370\pm{0.030}$ & $70.25\pm11.42$& -415.87&-1.36 && &$69.81\pm11.42$&$-0.433\pm0.080$&$0.85\pm0.38$ &$ -0.46\pm0.80$&-418.53&-1.19\\
    \hline
\end{tabular}}
\end{table*}

\begin{figure*}[ht!]
\centering
\plotone{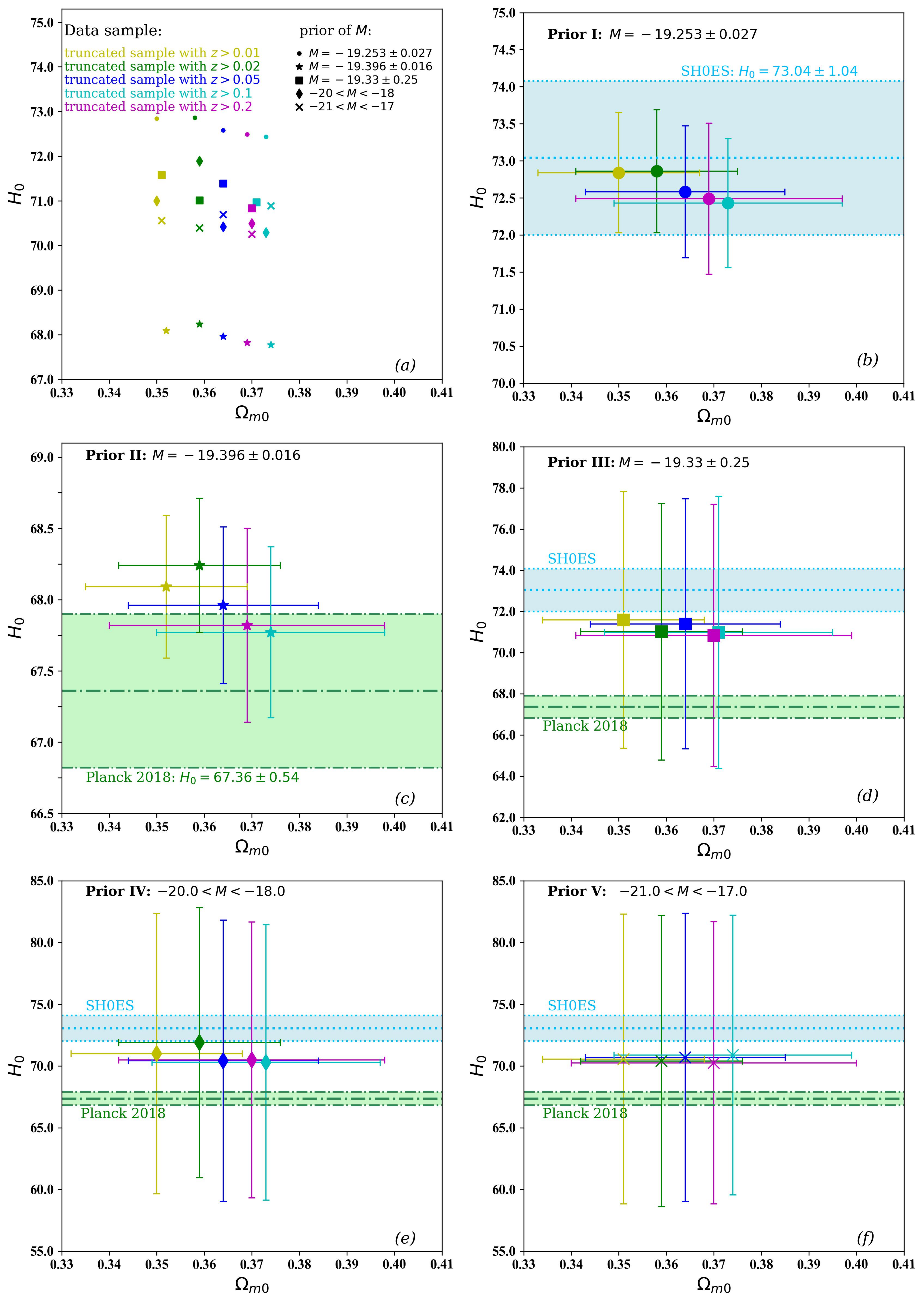}
\caption{The observational constraints on $\Omega_{\rm m0}$ and $H_0$. Panel (a) displays the mean values of $\Omega_{\rm m0}$ and $H_0$, and panels (b-f) show the mean parameter values and 68\% confidence level errors obtained for each of the five different priors of $M$. $H_0$ has units of km s$^{-1}$ Mpc$^{-1}$.} 
\label{fig:plot1}
\end{figure*}

The SNe Ia observational constraints on $\Omega_{\rm m0}$ and $H_0$ are presented in Table {\ref{tab:results}} and Figure {\ref{fig:plot1}}, including the results obtained from the truncated samples with five different redshift thresholds (Section {\ref{sec:truncated_redshift}}), for the five priors of $M$ (Section {\ref{sec:prior}}). In Table {\ref{tab:results}} we summarize the mean values of $\Omega_{\rm m0}$ and $H_0$ together with their 68\% confidence limits, where the five $M$ prior results from each (redshift-truncated) sample are grouped in the same vertical subpanel of the table. There are six panels in Figure {\ref{fig:plot1}}. The first panel displays the mean values of $\Omega_{\rm m0}$ and $H_0$ obtained from the five redshift-truncated samples, for the five priors of $M$. The other five panels show the mean values with marginalized 68\% confidence level errors of $\Omega_{\rm m0}$ and $H_0$ obtained for each prior of $M$.

We first focus on the effect of prior knowledge of $M$. The relevant information can be extracted from Table \ref{tab:results}. We study relative changes of the $\Omega_{\rm m0}$ and $H_0$ values when changing the prior on $M$, with the results obtained with \textbf{Prior I}, i.e., $M = -19.253\pm0.027$, taken as reference values. We see that relative differences in mean values of $\Omega_{\rm m0}$ are less than 1\% for the five redshift-truncated samples. Relative changes in mean values of $H_0$ are larger, ranging from $\sim$2\% to $\sim$7\%. The changes of the uncertainties of $\Omega_{\rm m0}$ are almost negligible. As expected, the uncertainty of $H_0$ is significantly positively related to the corresponding uncertainty of $M$. 

In passing, we note that the mean values of $\Omega_{\rm m0}$ obtained here are somewhat larger than most other estimates, e.g. \citet{2020A&A...641A...6P} find $\Omega_{\rm m0} = 0.3153 \pm 0.0073$ from CMB data, \citet{2023PhRvD.107f3522D} measure $\Omega_{\rm m0} = 0.3053 \pm 0.0050$ from CMB and non-CMB data, and \citet{2023PhRvD.107j3521C} find $\Omega_{\rm m0} = 0.313 \pm 0.012$ from non-CMB data, all in the flat $\Lambda$CDM model. This seems to be the case when SNe Ia data are used to constrain flat $\Lambda$CDM model parameters, for instance \citet{Rubinetal2023} find $\Omega_{\rm m0} = 0.356^{+0.028}_{-0.026}$ while \citet{DES2024} get  $\Omega_{\rm m0} = 0.352 \pm 0.017$.

We next investigate the effect of the truncation redshift used to eliminate lower-redshift SNe and so study the potential impact of the peculiar velocity subtraction. The relevant information can be extracted from Figure {\ref{fig:plot1}}. From panel (a) of Figure {\ref{fig:plot1}} we see two important trends: (i) independent of which redshift-truncated sample is used, mean values of $H_0$ obtained with \textbf{Prior I}, $M = -19.253\pm0.027$, are larger than those obtained using the other four priors; conversely, mean values of $H_0$ obtained with \textbf{Prior II}, $M = -19.396\pm0.016$, are obviously smaller than those obtained using the other four priors; (ii) under a given $M$ prior, different truncation redshifts result in only small relative changes in the mean values of $H_0$, less than 1\%, while relative changes in the mean values of $\Omega_{\rm m0}$ range from 0\% to $\sim$6\%.

Panel (b) of Figure {\ref{fig:plot1}} indicates that the estimates of $H_0$ obtained with $M$ \textbf{Prior I} are consistent with the SH0ES project \citep{2022ApJ...934L...7R} value, $H_0 = 73.04\pm1.04$ km s$^{-1}$ Mpc$^{-1}$, independent of the truncation redshift. In turn, panel (c) of Figure {\ref{fig:plot1}} indicates that the estimates of $H_0$ obtained with $M$ \textbf{Prior II} are consistent with the Planck 2018 \citep{2020A&A...641A...6P} result, $H_0 = 67.36 \pm 0.54$ km s$^{-1}$ Mpc$^{-1}$. Finally, panels (d-f) of Figure {\ref{fig:plot1}} show that the estimates of $H_0$ obtained with $M$ \textbf{Priors III--V} are consistent with both the Planck 2018 result and the SH0ES result. Additionally, from panels (b-f), we see that the estimates of $\Omega_{\rm m0}$ are consistent at 68\% confidence level, regardless of the chosen truncation redshift.

We also use Bayesian evidence \citep{rob95,2008ConPh..49...71T} to judge which prior on $M$ is preferred by the dataset. We use $\ln{B_i}$ to denote the natural logarithm of the Bayesian evidence for each scenario, and $\ln B_{i0}=\ln B_i-\ln B_0$ for the  relative log-Bayesian evidence, where ``0'' corresponds to the model with \textbf{Prior I}.  The model with the smallest $|\ln{B_i}|$ is the preferred model, and therefore we use the model with \textbf{Prior I} as the reference model in model comparison.  According to the empirical scale for evaluating the strength of evidence, $\rvert\ln{B_{i0}}\rvert \in (0, 1.0), (1.0, 2.5), (2.5, 5.0)$, and $(5.0, \infty)$ correspond to inconclusive, weak, moderate, and strong evidence, respectively \citep{2008ConPh..49...71T}. The corresponding values of $\ln{B_i}$ and $\ln{B_{i0}}$ are listed in Table \ref{tab:results}, where we notice $\rvert\ln{B_{i0}}\rvert < 2.0$ for all cases, which means that there is weak evidence that \textbf{Prior I} is preferred.

It is also necessary to test whether the overall findings are  model-independent. Hence, we repeat the above analysis by replacing the flat $\Lambda$CDM model with the cosmography model \citep{2002ApJ...569...18T, 2004CQGra..21.2603V,2006ApJ...649..563S,2011PhLB..702..114X}.  In the spatially flat scenario, the cosmography model parameter set is $\textbf{p} = (H_0, q_0, j_0, s_0)$, where $q_0$, $j_0$, and $s_0$ denote the present-day values of the deceleration, jerk, and snap parameters, respectively. The observational constraints on the cosmography model parameters, along with the Bayesian evidence in each case, are listed in the right part of Table \ref{tab:results}. The cosmography model results are very similar to the flat $\Lambda$CDM model results, which suggests that our results are robust and model independent.

\section{Conclusion}
We have investigated the impact of different data-moitvated priors for the peak absolute magnitude $M$ of SNe Ia on the measurement of the Hubble constant $H_0$ from apparent magnitude-redshift data from the Pantheon+ sample with 1701 light curves of 1550 distinct SNe Ia in the redshift range $0.001 < z < 2.26$. We use five distinct $M$ priors, discussed in the literature, and show that the choice of $M$ prior can significantly affect the measured $H_0$ value, but has marginal impact on the estimate of $\Omega_{\rm m0}$. The variations in mean $H_0$ values range between 2\% and 7\%, with uncertainty in $H_0$ positively correlated to $M$ prior uncertainty. Conversely, changes in the mean $\Omega_{\rm m0}$ values are less than 1\%, and changes in $\Omega_{\rm m0}$ uncertainties remain nearly negligible. 

We have also explored the influence of peculiar velocity subtraction by using five different low-$z$-truncated SNe Ia samples. The redshift threshold of the truncated sample can significantly impact the estimate of $\Omega_{\rm m0}$, while it has only a small effect on the derived $H_0$ value. 

The loose priors for $M$ yield $H_0$ values consistent with both Planck 2018 and SH0ES results at a 68\% confidence level. Considering the critical role of a reasonable and effective $M$ prior in obtaining reliable constraints on $H_0$, our study advocates the need for accurate estimates of the intrinsic dispersion of $M$ among SNe Ia, requiring a thorough understanding of contributing factors, including contributions from the diversity of SNe Ia progenitors \citep{2011NatCo...2..350H,2018PhR...736....1L,2023ApJ...953...13F}, and from correlations between SNe Ia  absolute magnitudes and host galaxy properties \citep{2020A&A...644A.176R,2023MNRAS.518.1985M}. Finally, the similarity between the cosmography model results and the $\Lambda$CDM model results indicate that our results are robust and model independent.

\begin{acknowledgments}
YC would like to thank Sheng Yang for helpful discussions on the observed SN Ia light curves.
This work has been supported by the National Key Research and Development Program of China (No.\ 2022YFA1602903), the National Natural Science Foundation of China (Nos.\ 11988101 and 12033008), the China Manned Space Project (No.\ CMS-CSST-2021-A01), and the K.\ C.\ Wong Education Foundation. SK gratefully acknowledges support from the Science and Engineering Research Board (SERB), Govt.\ of India (File No.~CRG/2021/004658).

\end{acknowledgments}

\vspace{5mm}

\bibliography{main}{}
\bibliographystyle{aasjournal}

\end{document}